\documentclass{rmf-d}
\usepackage{rmfbib,multicol,times,epsf,amsmath,amssymb,cite}
\usepackage[T1]{fontenc} 
\usepackage[]{caption2}
\usepackage{graphicx}
\usepackage{subfigure}
\usepackage{blindtext}
\usepackage[monochrome]{color}
\usepackage{hyperref}
\usepackage{minted}
\def\rmfcornisa{}
\def\rmfcintilla{}

\begin{document}
\pagestyle{plain}

%
%
\title{Microscopic dynamics of contagion using active Brownian particles: universal scaling and propagation controlled by protection}
%
%
\author{Isela Sicarú Regalado-Alvarado, and Francisco Alarcón\thanks{paco@fisica.ugto.mx}}
\address{División de Ciencias e Ingenierías, Universidad de Guanajuato, Lomas del Bosque 103, 37150 León, México.}
\maketitle
%
%
\recibido{\date{\today}}{??
\vspace{-12pt}}
\begin{abstract}
\vspace{1em} 
Understanding how individual protection and population density influence epidemic spreading remains a central challenge in epidemiology. 
While classical compartmental models successfully describe the temporal evolution of epidemics, they do not explicitly account for the microscopic motion and spatial organisation of individuals.
Here, we investigate contagion dynamics using an agent-based model of Active Brownian Particles (ABPs), where self-propelled agents interact through local contact and a prescribed fraction of the population is protected.
By systematically varying the protected-agent fraction and the population density, we identify two distinct contagion regimes separated by a crossover protection of approximately $30\%$.
Below this threshold, the maximum contagion rate follows the universal scaling $\nu_{\mathrm{max}}\propto\phi^{1/2}$, indicating that disease transmission is  governed primarily by  frequency of encounters. 
Above the threshold, universal scaling is lost and protection becomes the dominant mechanism controlling epidemic spreading, with the strongest suppression occurring in low-density populations.
These results demonstrate that microscopic active-matter models provide a powerful framework for investigating epidemic dynamics beyond the assumptions of well-mixed population models and reveal how spatial organisation and individual protection jointly determine contagion dynamics.

\vspace{1em}
\end{abstract}
\keys{ \bf{\textit{Active Brownian particles; Epidemic spreading; Agent-based simulations; Contagion dynamics; Universal scaling.
}} \vspace{-8pt}}

\begin{multicols}{2}

\section{Introduction}
 The appearance of diseases as seen during the SARS-CoV-2 outbreak. The number of infected cases increased, with no means available to prevent or treat the disease. COVID-19 appeared suddenly and being a highly contagious disease, there was little or no information available at that moment. This led to a lack of treatment and prevention protocols.
 
 Very recent research in animal fur production reveals the presence of viruses with zoonotic potential, particularly in high-density environments with limited sanitary measures \cite{zhao_2024_farmed}. The relevance of such investigation resides in the fact that the cross-species transmission of viruses from animals to humans drives infectious disease
emergence, previous works have been suggested to be the potential hosts for a variety of human viruses, including influenza A virus (IAV), and SARS-CoV-2 \cite{InfluenzA,Mink2Humans}.
Additionally, there are a couple of recent examples of pathologies that could compromise human health worldwide and cause an epidemic. For example the current case of the cruise ship with confirmed passengers with hantavirus \cite{hantavirus} or the recurring outbreaks of viral infections such as measles \cite{measles}.

The spread of all these infectious diseases can be analysed from a theoretical perspective using mathematical models of population growth, in which the population is generally divided into two classes according to their epidemiological status: susceptible and infected individuals \cite{SIFrancisca}.
 The first conventional analytical model is called the SI model (Susceptible - Infected).
 This model predicts the dynamics of contagion by following the dynamics of susceptible agents at time $t$ as $S(t)$ and the infected agents as $I(t)$. 
 Different aspects can be studied by introducing additional parameters. However, this results in much more complex differential equations, requiring advanced techniques and methodologies to solve them. 
 For example, the effect of social distancing \cite{Mallela} and lockdown \cite{NanaSIR}, even the effects of virus mutations that precede waves of contagion \cite{Waves}. While, other models focus on the hydrodynamics of viral transmission through saliva droplets \cite{Saliva}.
 

An alternative approach to study the dynamics of contagion in a population is the use of simulations with self-propelled agents.
The main advantage of these methods is that they provide information on how the spatial distribution and the velocity of the agents affect the contagion dynamics. 
In other words, the influence of \textit{microscopic} effects on the collective contagion dynamic.
Moreover, thanks to this microscopic approach, other parameters that are often neglected in conventional epidemiological approaches can be taken into account, for example the alignment of the agents' motion \cite{Huepe}, the confinement in hallways \cite{Pasillo}, or in more general terms, any feedback that modifies the epidemic spreading \cite{Levis}. 
Several groups around the world have investigated microscopic epidemiological models; one of the pioneer works appeared almost 20 years ago\cite{peruani_sibona_2008} and since then the community has tried to understand the contagion dynamics in terms of the typical features of the active agents like when MIPS emerge in a system of susceptible and infected agents \cite{MIPS_SIR_2022}.
One interesting approach is the one proposed by the group of Guzmán-Lastra, et al. \cite{Francisca}. They used Active Brownian Particles (ABPs) to represent walkers or agents that are able to change their internal state and reproduce the curves of the SIR model (Susceptible-Infected-Recovered). They determined the impact of contagious radius, particle velocity, and particle density as parameters that can promote virus spread.

Similarly, here in this work, we have used ABPs to model agents where some of them are either Susceptible or Infected, but we have also added Protected agents, in order to take in count vaccinated persons or with some kind of protection, like de facemasks in the case of COVID-19. 

\textcolor{red}{The objective of this work is to investigate how the inclusion of protected agents affects the microscopic contagion dynamics in a population of active Brownian particles. In particular, we aim to characterize how the maximum contagion rate depends on the population density and the fraction of protected agents, and to identify the regimes that emerge from their interplay.
}

\textcolor{red}{
We find two distinct regimes in terms of the maximum contagion rate: one in which the normalised rate is approximately independent of the fraction of protected agents, and another in which the contagion dynamics depend on both the protected fraction and the population density.
}
\section{Active Matter}
Active matter can be defined as materials made up of many interacting units, where each unit moves autonomously by transforming energy into motion. These systems are out of equilibrium. And manifest in both living and non-living worlds and in a wide range of scales \cite{Review_2016,Metareview2025}.
There is a plethora of models for active matter, but historically, two main minimal models can be identified in the context of numerical simulations of active particles. On the one hand, the Vicsek model, which takes in count self propulsion and effective alignment among the neighbours.
This model proposed by Vicsek, was used to describe flocks of birds and schools of fish \cite{vicsek_model}, it can be seen as a very raw approximation of the hydrodynamic interactions between active particles.
On the other hand, we have a model of self-propelled motion in a dissipative environment.  The centre of mass and the direction of the propulsive force are subjected to white noise, which introduces a diffusive component to the overall dynamics. It is the so-called Active Brownian Particles (ABPs), where passive Brownian dynamic equations of motion have been modified to describe the position and the direction of particles \cite{MIPS_paper}. With applications typically at the microscale, to understand bacterial motion, active colloids, etc.


\subsection{Passive Brownian dynamics}
Brownian motion was first observed by Robert Brown while examining pollen grains suspended in water. He noticed that the grains exhibited persistent irregular motion \cite{AboutRobertBrown}. Later, Albert Einstein explained that this motion originates from collisions with the surrounding fluid molecules and derived the diffusion law for suspended particles \cite{PaperEinstein1905}.

Shortly afterward, Langevin introduced a stochastic description of Brownian motion by incorporating a random force into Newton's equation of motion. This approach led to the Langevin equation, 
\begin{equation}
m \ddot{r} = - \gamma \dot{r} + \omega \xi(t) + F(r),
\label{lang_eq}
\end{equation}
where $\gamma$ is the friction coefficient due to viscosity which depends on velocity and $\omega$ is the coefficient of the fluctuating term. The stochastic term $\xi(t)$ has the following properties: $ \langle \xi_i (t) \rangle = 0$ and $\langle \xi_i (t_1) \xi_j (t_2) \textcolor{red}{\rangle} = \delta_{ij} \delta(t_2-t_1)$.

In modern terms, this random force is modelled as Gaussian white noise, leading to the stochastic differential equation given by equation \eqref{lang_eq}, which describes the dynamics of a Brownian particle under the action of an external force $F(\mathbf{r})$  \cite{Langevin1908}.

If we consider a system of particles that do not interact  with each other ($ F(r)=0$) and a low Reynold’s number regime, in other words, the overdamped limit where $|\gamma \dot{r} | >> |m \ddot{r}|$, the Langevin equation \eqref{lang_eq} simplifies to $\gamma \dot{r} = \omega \xi (t)$. 
By assuming that the coefficients  $\omega$ and $\gamma$ are spatially homogeneous and using stochastic calculus tools, it is possible to identify the Einstein diffusion coefficient \cite{Mazo2002} as  
\begin{equation}
D = \frac{\omega ^2}{2 \gamma ^2}.
\label{Coef_Dif_Einstein}
\end{equation}

Now, if we consider the presence of a conservative force field $F(r) \neq 0$ in the expression \eqref{lang_eq} at the inertial limit, it is possible to derive the Smoluchowski Diffusion equation \cite{Mazo2002}, by assuming a Boltzmann distribution for the random positions of the Brownian particles, we get the so-called Fluctuation-Dissipation theorem where the temperature is related with the amplitude of both the stochastic force and the friction coefficient as
\begin{equation}
\omega ^2 = 2 k_B T \gamma.
\label{cond-S_eq}
\end{equation}

Then we can rewrite \eqref{lang_eq} in terms of the macroscopic parameters as
\begin{equation}
\dot{r} = \sqrt{2D} \xi(t) + \frac{D}{k_B T} F(r)
\label{overdamped_Eq}
\end{equation}
in the overdamped limit.


\subsection{Active Brownian dynamics}
If activity is introduced into Passive Brownian dynamics, then it can describe particles that are capable of transforming their internal energy into motion. These particles can represent microorganisms such as bacteria; moreover, this is not limited to objects on the microscopic scale. Using ABPs allows us to describe macroscopic events like a swarm, a bank of fish, and even pedestrians\cite{Review_2016}. The Langevin equation, shown in equation \eqref{L_sim}, describes the motion of the agents with a constant-magnitude force $F_A$, which represents the activation force of the agents due to the transformation of internal energy into motion. 

\begin{equation}
m \ddot{\mathbf{r}}_i = - \gamma \dot{\mathbf{r}}_i+ \omega \mathbf{\xi}_i + \sum _{j\neq i} \mathbf{F}_{ij}    + \mathbf{F}_A,
\label{L_sim}
\end{equation} 
where the force $\mathbf{F}_{ij} $ does not describe an external force; instead, it describes the interaction between particles. 
The force $\mathbf{F}_{ij} = -\nabla V_{WCA}(r_{ij}) $ describes the repulsive interactions between ABPs ruled by the Weeks-Chandler-Andersen (WCA) potential presented in equation \eqref{WCA_potential}
\begin{equation}
    V_{WCA}(r_{ij}) = 4 \epsilon \Big[ \Big( \frac{\sigma}{r_{ij}} \Big)^{12} - \Big( \frac{\sigma}{r_{ij}} \Big)^6 \Big] + \epsilon.
    \label{WCA_potential}
\end{equation}
Here, $\sigma$ can be seen as the diameter of the particle, $\epsilon$ determines the interaction strength, and $r_{ij}$ is the separation centre-to-centre between two particles. With an upper limit of cut-off at $r=2^{1/6} \sigma$, beyond which $V_{WCA}=0$\cite{ABP_Joakin2014}.

In the overdamped limit and by applying the Einstein-Smoluchowski diffusion relations, the expression \eqref{L_sim} simplifies to  
\begin{equation}
\dot{r} = \sqrt{2D} \xi_i(t) + \frac{D}{k_B T} \sum _{j\neq i} \mathbf{F}_{ij}    + v_0 \hat{e}_i.
\label{overdamped_ABP}
\end{equation}

\textcolor{red}{All simulations are performed in reduced units. The particle diameter $\sigma$ defines the characteristic length scale, while the WCA energy parameter $\epsilon$ and the thermal energy $k_B T$ define the energy scale. We set
\begin{equation}
\sigma=\epsilon=k_B T=D=1,
\end{equation}
where $D$ is the translational diffusion coefficient. The corresponding Brownian diffusion time, $\tau_B=\sigma^2/D$, therefore defines the reduced time unit. 
Within this convention, forces are expressed in units of $k_B T/\sigma$. Furthermore, the rotational diffusion coefficient is not an independent parameter, since $D_R=3D/\sigma^2$, and the self-propulsion speed is given by $v_0=F_A D/(k_B T)$. 
Thus, in the reduced units employed here, $D_R=3$ and $v_0=F_A$. The integration time step is $\Delta t=10^{-6}$.
No direct mapping to physical units is assumed, as the model is intended to provide a generic microscopic description of contagion dynamics.}

We obtain an equation that describes the position of each agent. 
\begin{equation}
\begin{split}
\mathbf{r}_i(t+\Delta t) ={}& \mathbf{r}_i(t)
 + \sqrt{2D\Delta t}\,\xi_i(t) \\
&+ \frac{D}{k_B T}\sum_{j\neq i}\mathbf{F}_{ij}\Delta t
 + v_0\hat{e}_i\Delta t .
\end{split}
\label{positionone}
\end{equation}
the iterative equation \eqref{positionone} follows the computational implementation developed in Ref. \cite{ermark1978}.

\section{Model}
The dynamics of the agents is described considering $N$ particles of radius $\sigma/2$ moving in a two-dimensional square domain of size $L_x \times L_y$ with periodic boundary conditions in both the $x$ and $y$ directions and $L_x=L_y$. The particle motion follows the well-known active Brownian particle (ABP) dynamics described previously in equation \eqref{positionone} with the same self-propulsion speed $v_0$ for all agents. 

\begin{figure}[H]
\begin{center}
\includegraphics[width=\linewidth]{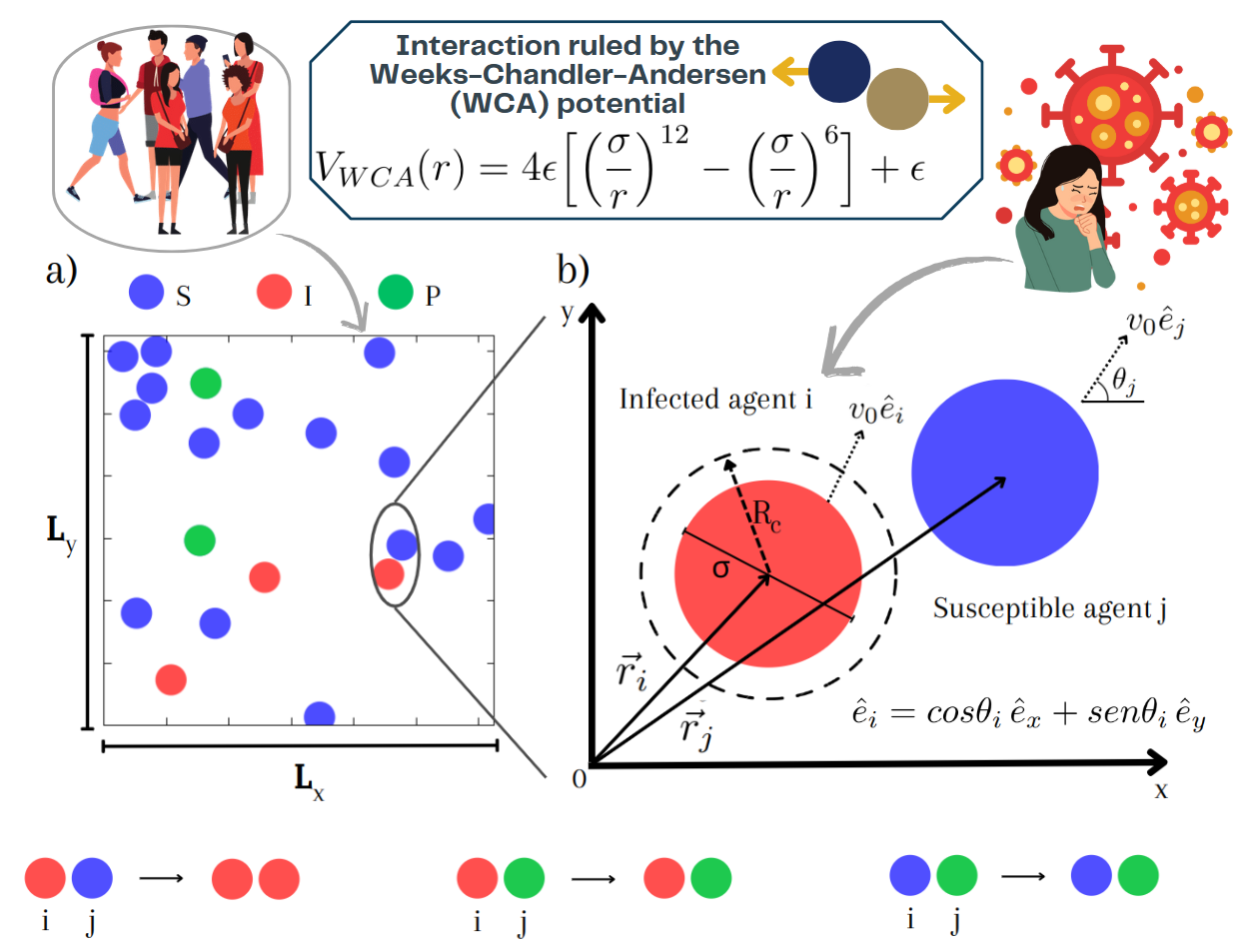} 
\caption{a) Sketch of the simulation box: $N$ moving agents of diameter $\sigma$ representing pedestrians. In a simulation box with periodic boundary conditions and the pair-interaction defined by the WCA potential. b) If an infected particle $i$ is at a distance $|r_i - r_j| \leq R_c$ of a susceptible agent $j$, it goes from susceptible to infected. The tree states: susceptible S (blue), infected I (red) and protected P (green) such that $S + I + P = N$. Image inspired by \cite{Francisca}.}
\label{model_image}
\end{center}
\end{figure}

For the contagion model, particles are divided into three compartments: susceptible, $S(t)$, infected, $I(t)$ and protected $P$, satisfying the conservation condition $S(t) + I(t) + P = N$. 
Only susceptible individuals can transition to the infected state according to the following contagion rule: a susceptible agent at $r_j$ will become infected if it is at a distance $|r_i - r_j| \leq R_c$ from an infected agent at $r_i$ for a certain time $T_c$. Where $R_c$ is the contagion radius measured from centre to centre and $T_c$ the contagion time, which is based on the characteristic infection time introduced by Reichhardt’s group in Ref.~\cite{MIPS_SIR_2022}. The parameter $T_c$  represents the effective interaction time required for transmission, expressed in terms of the numerical integration time step. In all simulations, we set $T_c = 10,000~\Delta t$, where $\Delta t$ denotes the integration time step of the equations of motion \eqref{positionone}.
Protected agents cannot change their internal state.

In Figure \ref{model_image} a) is a representation of our model. A simulation box of size $L_x \times L_y$ with periodic boundary conditions, the ABPs represent pedestrians in an open space, blue particles are susceptible, infected are in red and protected in green. In all simulations, initial positions $r_i$ and orientations $\hat{e}_i$ are randomly assigned. Likewise, in Figure \ref{model_image} b) is a more detailed scheme of the activity orientation and the range of infection is depicted. 

The simulations were carried out using an in-house code written in the C programming language \cite{CodeC,numericalrec}. The input parameters include the number of agents ($N=3000$) fixed in all the simulations and constrained to $N = S(t) + I(t) + P$, the packing fraction $\phi$ is changing by changing the size of the box, the active force ($F_A=50$) fixed in all the simulations, where $v_0 = DF_A/k_BT$ with $D/\textcolor{red}{k}_B T = 1$, the initial number of infected agents ($I(0)=1$) and the fraction of protected agents $\chi = P/N$. For all simulations, the number of agents is the same. Therefore, the protected ratio will change the number of initial susceptible agents $S(0)$. 

\textcolor{red}{The simulations were performed using high-performance computing resources provided by LANCAD and SECIHTI through the Miztli supercomputer at DGTIC-UNAM, and by the Laboratorio de Supercómputo del Bajío at CIMAT. These facilities were used to perform the large number of simulations required to systematically explore the parameter space.}

\section{Results}

The dynamic of infected agents $I(\tau)$ is the number of infected at each dimensionless time $\tau = it_\#\Delta t /t_S$. The term $it_\#$ is the iteration number and  
\textcolor{red}{$t_S=\frac{\sigma}{v_0}$, defined as the characteristic time required for an isolated active particle moving at the self-propulsion speed $v_0$ to travel a distance equal to its own diameter $\sigma$. It is called the self-propulsion time \cite{Ginot2018,Alarcon2017}.} 
The number of infected agents is calculated by averaging ten simulations for each protected ratio $\chi\in [0.0, 0.5 ]$, where $\chi = P / N$. 

\begin{figure}[H]
\begin{center}
\includegraphics[width=\linewidth]{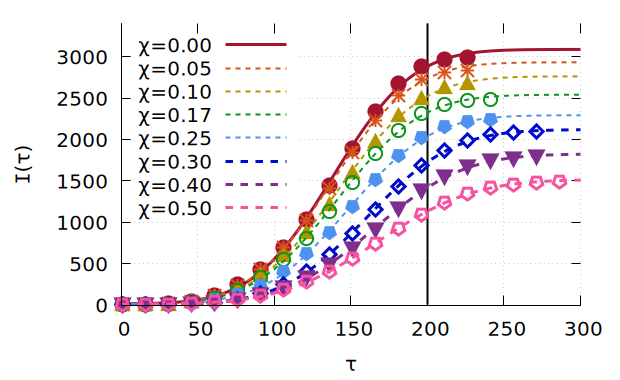} 
\caption{Plot of the contagion dynamics. The symbols represent the agent-based simulations, with a mobility force $F_A = 50$, $\phi = 0.056$. On the x-axis the dimensionless time $\tau$ and the y-axis the number of infected agents $I(\tau)$. Each coloured curve represents a protected ratio $\chi$. The continuous lines are the fitted erf functions.  The Data availability section \ref{Supp_info} leads to example clips from the results obtained of our simulations.}
\label{results_infected}
\end{center}
\end{figure}

Figure \ref{results_infected} shows the temporal evolution of the number of infected individuals, $I(\tau)$, until a steady state is reached. This steady-state value depends on the protected-agent fraction, $\chi$, and corresponds to the stage at which the number of susceptible agents becomes zero or nearly zero. As expected, increasing the fraction of protected agents reduces the final number of infected individuals.

The red circles on Figure \ref{results_infected} are the contagion dynamics with no protected agents ($\chi=0$) and it can be seen as a reference case to determine the impact of the protected ratio. 
We can observe the evolution of the infected agents from 1 to 3000 infected agents at $\tau \approx 200$.  
When there are protected agents, the evolution saturates also around $\tau \approx 200$ for $\chi < 0.3$, but once the protected ratio reaches $0.3$ (blue diamonds in Figure \ref{results_infected}) the saturation time shifts to longer times.

Additionally, the simulation data were fitted using a nonlinear least-squares Levenberg--Marquardt algorithm \cite{numericalrec}. The solid curves in Figure~\ref{results_infected} correspond to the resulting fits \cite{gnuplot}. To describe the temporal evolution of the number of infected individuals, we employed the error function, whose sigmoidal shape is equivalent to that of the logistic function, the analytical solution of the classical SI epidemic model. Therefore, the error function provides a robust empirical representation of the contagion curves while preserving the characteristic behaviour predicted by the SI model. The fitting function is given by
\begin{equation}
I(t)=A\,\mathrm{erf}\!\left[B(t-C)\right]+D,
\label{erf_funct}
\end{equation}
where the fitting parameters $A$, $B$, $C$, and $D$ are determined from the simulation data.

\begin{table*}[ht]
\centering
\begin{tabular}{cccccc}
\hline
\multicolumn{1}{|c|}{$\chi$} & \multicolumn{1}{c|}{A} & 
\multicolumn{1}{c|}{B} & \multicolumn{1}{c|}{C} & 
\multicolumn{1}{c|}{D} & \multicolumn{1}{c|}{$R^2$} \\ \hline
0.00 & 1538.08 & 0.0172 & 138.212 & 1551.86 & 0.99942 \\
0.05 & 1463.29 & 0.0169 & 136.993 & 1470.78 & 0.99967 \\
0.10 & 1380.13 & 0.0163 & 141.395 & 1384.55 & 0.99974 \\
0.17 & 1268.31 & 0.0165 & 141.156 & 1274.53 & 0.99970 \\
0.25 & 1140.65 & 0.0166 & 148.095 & 1153.77 & 0.99962 \\
0.30 & 1055.14 & 0.0161 & 160.731 & 1065.11 & 0.99983 \\
0.40 & 907.95 & 0.0153 & 164.922 & 917.94 & 0.99973 \\
0.50 & 752.81 & 0.0143 & 167.252 & 763.38 & 0.99978 \\
\hline
\end{tabular}
\caption{Parameters obtained from the nonlinear least-squares fits of the error function to the simulation data shown in Figure~\ref{results_infected} for different protected fractions $\chi$. The corresponding coefficient of determination, $R^2$, is also reported as a measure of the goodness-of-fit.}
\label{Table_parameters}
\end{table*}

\textcolor{red}{The quality of the fits was quantitatively assessed through the coefficient of determination, $R^2$. As shown in Table~\ref{Table_parameters}, the values of $R^2$ range from $0.9994$ to $0.9998$ for all protected fractions considered, indicating an excellent agreement between the fitted profiles and the simulation data. Moreover, no systematic deterioration or improvement of the goodness-of-fit is observed as the protected fraction increases. Thus, the error-function form provides a consistent description of the contagion curves over the entire range of $\chi$ investigated. Although the quality of the fits remains essentially unchanged, the fitting parameters exhibit a systematic evolution with the protected fraction. In particular, $A$ and $D$ decrease as $\chi$ increases, while $C$ generally increases and $B$ shows an overall decreasing trend. Since $\nu_{\max}=2AB/\sqrt{\pi}$ and $\tau_{\max}=C$, these changes are directly related to the reduction of the maximum contagion rate and the delay in the time at which it is reached as the protected fraction increases.}

\begin{figure}[H]
\begin{center}
\includegraphics[width=\linewidth]{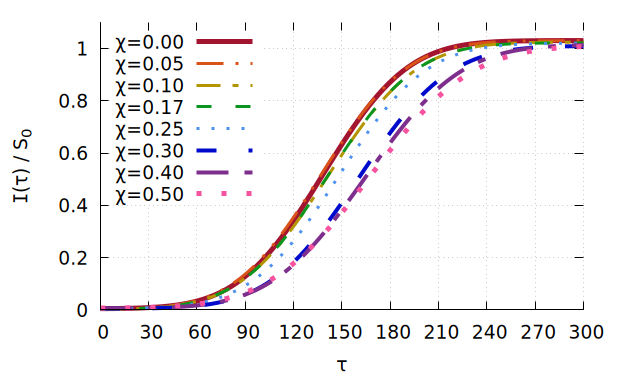} 
\caption{Normalised plot of the Contagion dynamics. The number of infected agents $I(\tau)$ divided by the initial number of susceptible agents $S_0$. On the x-axis the dimensionless time $\tau$.}
\label{results_normal_infected}
\end{center}
\end{figure}

To perform a detailed analysis and quantify the effect of protection,   Figure~\ref{results_normal_infected} shows the number of infected agents, $I(\tau)$, normalised by the initial number of susceptible agents, $S_0$, for different values of the protected-agent fraction, $\chi$. Each coloured curve corresponds to a different value of $\chi$. Remarkably, all curves with $\chi < 0.3$ collapse onto a single master curve, indicating that contagion dynamics is independent of the protected-agent fraction within this range. In contrast, for $\chi \textcolor{red}{\ge 0.3}$, the curves deviate progressively from the master curve, revealing a slower contagion dynamics as the fraction of protected agents increases. 
As we can observe from the blue, purple, and pink curves in Figure~\ref{results_normal_infected}. For clarity, only the fitted curves are shown.

 Although some values of $\chi \neq 0$ converge with the reference case, it can be seen that with values
 \textcolor{red}{with values starting at 30\%}
 of the agents protected, the normalised number of infected agents does not converge to the reference case. These values have an effect by modifying the contagion dynamics as they have a different slope compared to the reference case. The speed of contagion was analysed to explore in more detail the change in the contagion dynamics. The speed or evolution of contagion is the temporal derivative of  the infected agents,
 \begin{equation}
  I'(\tau) = \frac{d I(\tau)}{d \tau}.   
 \end{equation}

Since evolution of the number of infected agents $I'(\tau)$ follows the erf function, the expression for the speed of contagion follows a Gaussian \cite{Andrews1998}: 
\begin{equation}
I'(\tau) = \frac{2 A B}{\sqrt{\pi}} e^{-(B( \tau - C))^2}
\end{equation}

 Thus, we can define $\tau_{\mathrm{max}} =C$, where $\tau_{\mathrm{max}}$ is the time at which the maximum speed of contagion $\nu_{\mathrm{max}}$ is reached with $\nu_{\mathrm{max}} = (2 A B)(\pi)^{-1/2}$.
 In this context of contagion dynamics, these values are related to the saturation of health care infrastructures. They indicate the time when the speed of contagion is at its highest, that is, when the largest number of susceptible agents simultaneously become infected.

\begin{figure}[H]
\begin{center}
\includegraphics[width=\linewidth]{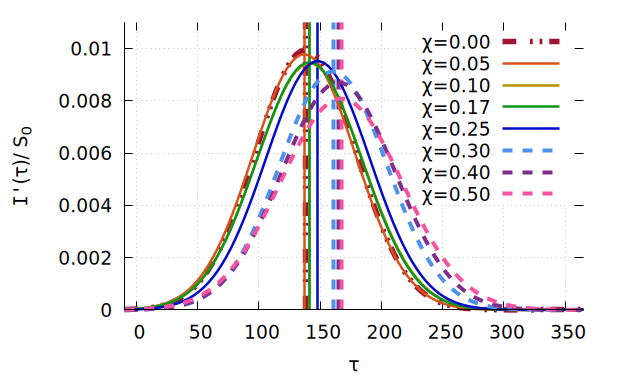} 
\caption{Normalised contagion rate as a function of time for different protected-agent fractions, $\chi$. The corresponding values of $\tau_{\mathrm{max}}$ are indicated by the vertical lines and denote the time at which the contagion rate reaches its maximum value, $\nu_{\mathrm{max}}$.}
\label{der_infected}
\end{center}
\end{figure}

The normalised contagion rate, $I'(\tau)/S_0$, is shown in Fig.~\ref{der_infected}. 
The Gaussian-like profiles provide a clear visualisation of the infection dynamics and reveal two main features. 

\textcolor{red}{The inflection point of the contagion curves shown in Figures~\ref{results_infected} and \ref{results_normal_infected} is directly related to the maximum contagion rate presented in Figure~\ref{der_infected}. For the error-function fit (equation (\ref{erf_funct}) the derivative reaches its maximum at $\tau_{\max}=C$. Thus, the inflection point, corresponding to the maximum slope of the contagion curve, coincides with the maximum of the contagion rate. The shift of this point toward larger times with increasing protected fraction is therefore consistent with the increase of $\tau_{\max}$ observed in Figure~\ref{der_infected}.}

First, the time required to reach the maximum speed of contagion, $\tau_{\mathrm{max}}$, and how it increases with the protected-agent fraction, $\chi$.
For low protection levels ($\chi\textcolor{red}{< 0.3}$), all curves reach their maximum at nearly the same time. This indicates that protection has little influence on the onset of the fastest spreading regime. The values of $\tau_{\mathrm{max}}$ are indicated by the corresponding vertical solid and dashed lines.

Second, the peak of the speed of contagion remains nearly unchanged for $\chi \textcolor{red}{< 0.3}$, whereas for larger protection levels ($\chi \textcolor{red}{\ge 0.3}$) the peak decreases and shifts to later times. These results suggest that increasing the percentage of  protected agents not only delays the progression of the epidemic, but also reduces its maximum speed of contagion.


So far, we have considered a fixed population density, corresponding to a packing fraction of $\phi = 0.056$. 
Next, we examine the influence of population density on the contagion dynamics in the presence of protected agents.
The packing fraction, $\phi$, was varied by changing the size of the square simulation box while keeping the total number of agents fixed at $N = 3000$. The dependence of the maximum speed of contagion on the density of the agents is examined.

\begin{figure}[H]
\begin{center}
\includegraphics[width=\linewidth]{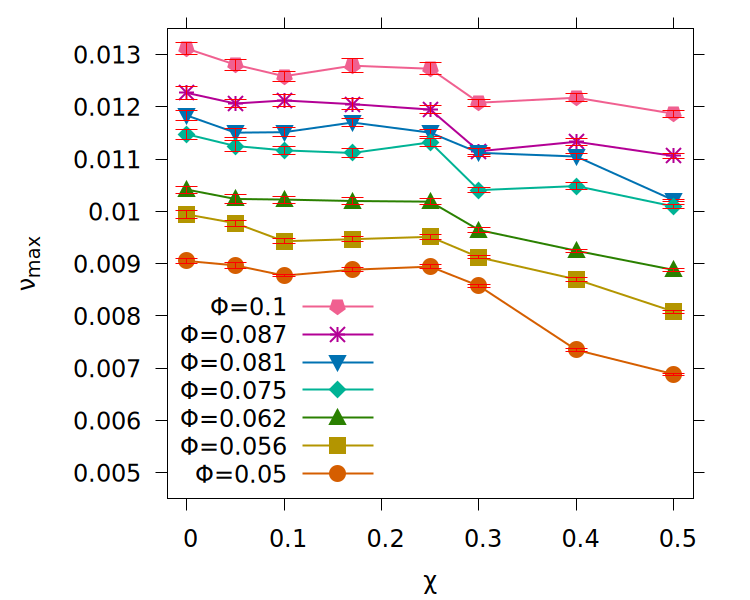} 
\caption{Maximum speed of contagion, $\nu_{\mathrm{max}}$, as a function of the protected-agent fraction, $\chi$, for different packing fractions, $\phi$. Different packing fractions are represented by coloured symbols connected by solid lines to guide the eye.}
\label{BigPicture_speed}
\end{center}
\end{figure}

Figure~\ref{BigPicture_speed} shows the maximum speed of contagion, $\nu_{\mathrm{max}}$, as a function of the protected-agent fraction, $\chi$, for different packing fractions. The curves of the maximum speed of contagion, $\nu_{\mathrm{max}}$, do not intersect for any packing fraction. Nevertheless, they exhibit the same qualitative dependence on the protected fraction, $\chi$. 
Specifically, $\nu_{\mathrm{max}}$ remains nearly constant for $\chi < 30\%$, defining an initial plateau, and then decreases monotonically as the fraction of protected agents increases. 
Both the plateau value and the subsequent decay depend on the packing fraction.
Higher packing fractions result in larger plateau values and a weaker reduction of the maximum contagion rate at high protection levels, whereas lower packing fractions exhibit lower initial contagion rates followed by a more pronounced decrease as $\chi$ increases. 

The corresponding contagion dynamics, normalised contagion rates, and temporal evolution of the contagion rate for the remaining packing fractions are presented in Appendix~\ref{Appenix:1}.

\begin{figure}[H]
\begin{center}
\includegraphics[width=\linewidth]{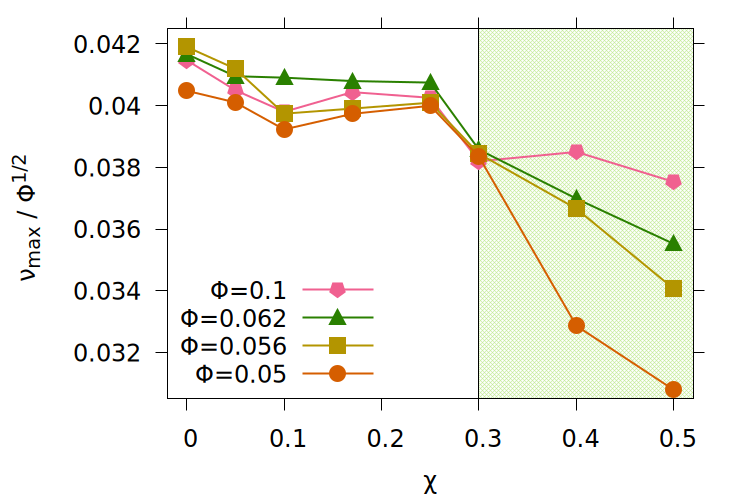} 
\caption{Normalised maximum contagion rate, $\nu_{\mathrm{max}}\phi^{-1/2}$, as a function of the protected-agent fraction, $\chi$, for different packing fractions, $\phi$. Coloured symbols correspond to different values of $\phi$, and the solid lines are included as guides to the eye. The shaded region highlights the transition at approximately $\chi = 30\%$.}
\label{Speed_trim}
\end{center}
\end{figure}

We normalised the maximum contagion rate by $\phi^{1/2}$ to further investigate the role of population density, as shown in Figure~\ref{Speed_trim}.

\textcolor{red}{A crossover between the two contagion regimes is observed around $\chi=0.30$ in Figure~\ref{Speed_trim}. Below this value, the normalized maximum contagion rate follows an approximately universal behavior, whereas for larger protected-agent fractions the curves progressively deviate from this scaling and the effect of protection becomes the dominant mechanism controlling the dynamics of the epidemic. Given the discrete values of $\chi$ considered in our simulations, $\chi=0.30$ should be regarded as a reference value for the crossover rather than as a precisely determined critical point.}

A remarkable collapse of all curves is observed for $\chi < 0.3$, indicating that, in the low-protection regime, the maximum speed of contagion scales as
\begin{equation}
    \nu_{\mathrm{max}} \propto \phi^{1/2}.
\end{equation}
\noindent
\textcolor{red}{This scaling dependence can be interpreted in terms of the microscopic encounter dynamics of the agents. In the low-protection regime, most encounters involving infected agents can occur with susceptible individuals, and the contagion dynamics are therefore primarily controlled by the frequency of effective susceptible--infected encounters.
In two dimensions, the characteristic interparticle separation scales as $\ell\sim\rho^{-1/2}$, where $\rho$ is the surface number density. For a fixed characteristic self-propulsion velocity, the corresponding encounter timescale can be estimated as $t_{\mathrm{enc}}\sim\ell/v_0$, suggesting an encounter frequency proportional to $\rho^{1/2}$. Since $\rho\propto\phi$ in the present system, this simple geometrical argument is consistent with the sublinear dependence $\nu_{\mathrm{max}}\sim\phi^{1/2}$ observed in our simulations. A nonlinear dependence of contact rates on population density has also been reported in spatial epidemic models by Hu \textit{et al.}~\cite{Hu2013}, who showed that contact rates increase with population density but need not follow a simple linear dependence over the entire density range. As the protected-agent fraction increases, however, the number of effective susceptible--infected encounters is progressively reduced, and population density alone no longer controls the contagion rate. This provides a possible microscopic explanation for the departure from the $\phi^{1/2}$ scaling observed at larger values of $\chi$.}

Moreover, the departure from the universal curve depends systematically on the packing fraction. Lower packing fractions exhibit a stronger reduction in the normalised contagion rate than denser systems, whereas the highest-density systems remain comparatively less affected. These results suggest that high population densities partially compensate for the reduction in contagion caused by protected individuals. Where the higher frequency of encounters between agents still promotes disease transmission. 
Conversely, at low densities, where encounters are intrinsically less frequent, increasing the protected-agent fraction leads to a much more pronounced suppression of contagion.

For completeness, Appendix~\ref{Appendix:2} shows the normalised maximum contagion rate, $\nu_{\mathrm{max}}\phi^{-1/2}$, as a function of the protected-agent fraction, $\chi$, for all packing fractions analysed in this work.

A clear crossover is observed at approximately $\chi \approx 0.3$, in Figures \ref{Speed_trim} and \ref{All_Speed_trim}, separating two distinct contagion regimes. Below this threshold, the maximum contagion rate obeys the universal scaling $\nu_{\mathrm{max}} \propto \phi^{1/2}$. Above this threshold, the curves progressively deviate from the universal scaling, and the protective effect becomes dominant. The protective effect progressively reduces the maximum contagion rate, with the reduction becoming more pronounced as the packing fraction decreases.

\section{Conclusion}
In this work, we proposed a microscopic epidemiological model based on Active Brownian Particles to investigate the influence of individual protection and population density on contagion dynamics. Unlike classical compartmental models, the present approach explicitly incorporates the motion and spatial distribution of individuals. This enables the study of microscopic mechanisms governing the transmission of the disease.

The simulations reveal the existence of two distinct contagion regimes separated by a \textcolor{red}{protected-agent fraction associated with the crossover} of approximately $\chi \approx 0.3$. Below this threshold, the maximum speed of contagion is practically independent of the fraction of protected agents and follows the universal scaling
$\nu_{\mathrm{max}} \propto \phi^{1/2}$. This indicates that the contagion dynamics is primarily controlled by the frequency of encounters between individuals.
Above this threshold, the normalised maximum contagion rate progressively deviates from this universal scaling. This demonstrates that protection becomes the dominant mechanism controlling the spread of the disease.
In this regime, the suppression of the speed of contagion is considerably stronger in dilute populations than in dense ones.

These findings show that microscopic agent-based models provide a powerful framework for investigating epidemic spreading beyond the assumptions of well-mixed population models.  The model can explicitly account for mobility and local interactions. It  captures how spatial organisation and population density influence the effectiveness of protective measures. Which reveals behaviours that cannot be directly inferred from conventional SI-type models.

\textcolor{red}{Future work will focus on a more detailed characterization of the crossover around $\chi\approx0.30$ by exploring additional values of the protected-agent fraction in this region. This will allow us to determine more precisely the location and width of the crossover and to investigate whether it is associated with a genuine critical phenomenon.} 
\textcolor{red}{In addition, we plan to compare the present results with Epidemic Percolation Network models \cite{Kenah_2011,Yang_2015,Hiraoka_2022} 
to establish a quantitative relationship between the spatial organization of agents and the emergence of
large-scale contagion.
Such an analysis may also provide further insight into the microscopic origin of the density scaling observed in the low-protection regime by relating the effective connectivity
of contagion pathways to the packing fraction and protected-agent
fraction.}
\textcolor{red}{
\section*{Data availability} \label{Supp_info}
Supplementary information, including simulation videos and related files, is available at: 
\href{https://github.com/SicaReAl/Contagion_Dynamics_ABPS_Supplemental_document/tree/main}{Supplemental$\_$document}}

\textcolor{red}{
The data that support the findings of this study are available
from the corresponding author upon reasonable request.}

\section*{Acknowledgements} 
 I.S.R.-A would like to thank SECIHTI through Grant CVU: 2058169.
This work was possible thanks to the access to HPC-time granted by the following institutions: a) LANCAD and SECIHTI on the supercomputer \href{http://www.lancad.mx/?p=59}{Miztli} at DGTIC UNAM, calls 2024-2026.  b) \href{https://supercomputo.cimat.mx/inicio}{Laboratorio de Supercómputo del Bajío} CIMAT through project  "Supercómputo como motor de colaboraciones academia-industria en conjunto con el Instituto de Innovación, Ciencia y Emprendimiento para la Competitividad para el Estado de Guanajuato (IDEA GTO)".
We thank Arturo Jimenez, Jesús Bernal, Renato Iturriaga, Ana Laura Benavides and Nana Geraldine for helpful discussions.

\end{multicols}
\appendix
\renewcommand{\thesection}{\Roman{section}}
\section{Contagion Dynamics for Additional Packing Fractions} \label{Appenix:1}
This appendix presents the complete set of results for the additional packing fractions analysed in this work, namely $\phi=\{0.05,\,0.062,\,0.075,\,0.0813,\,0.0875,\,0.10\}$. For each packing fraction, we show the contagion dynamics, $I(\tau)$, the normalised contagion dynamics, $I(\tau)/S_0$, and the normalised contagion rate, $I'(\tau)/S_0$.

The symbols correspond to the averages obtained from ten independent agent-based simulations for each protected-agent fraction, $\chi$, using a fixed active force of $F_A=50$. The solid curves represent the corresponding fits described in the main text, while different colours denote different values of the protected-agent fraction. In all cases, the horizontal axis corresponds to the dimensionless time, $\tau$.

\begin{figure}[ht]
\centering
\begin{minipage}{0.3\textwidth}
    \centering
    \includegraphics[width=\linewidth]{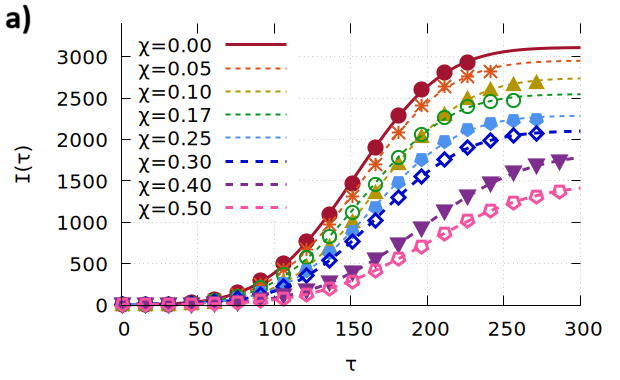}
\end{minipage}
\hfill
\begin{minipage}{0.3\textwidth}
    \centering
    \includegraphics[width=\linewidth]{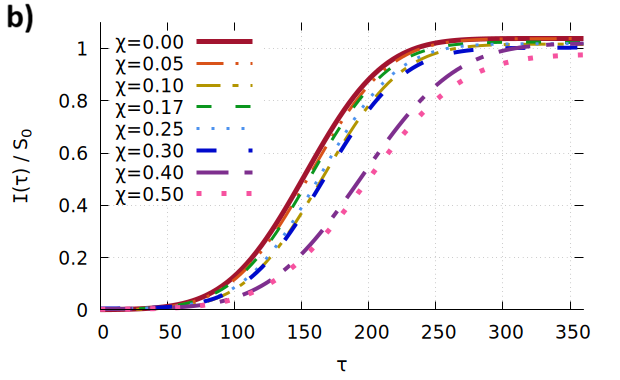}
\end{minipage}
\hfill
\begin{minipage}{0.3\textwidth}
    \centering 
    \includegraphics[width=\linewidth]{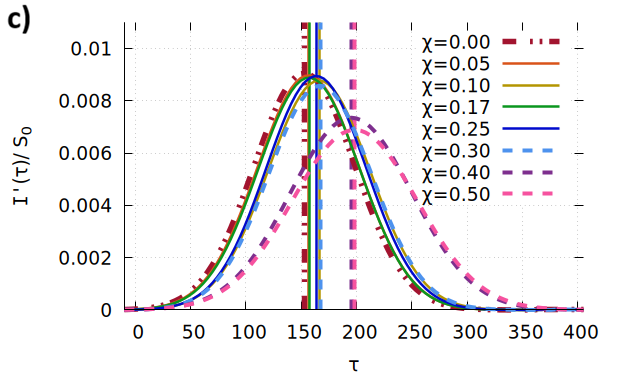}
\end{minipage}
\caption{\textbf{Results of the Packing fraction of 0.05} (a) Contagion dynamics, $I(\tau)$; (b) normalised contagion dynamics, $I(\tau)/S_0$; and (c) normalised contagion rate, $I'(\tau)/S_0$. Symbols represent the averages of ten independent agent-based simulations, while the solid curves correspond to the fitted functions. Vertical lines in panel (c) indicate $\tau_{\mathrm{max}}$, the time at which the maximum contagion speed, $\nu_{\mathrm{max}}$, is reached.}
\end{figure}

\begin{figure}[ht]
\centering
\begin{minipage}{0.3\textwidth}
    \centering
    \includegraphics[width=\linewidth]{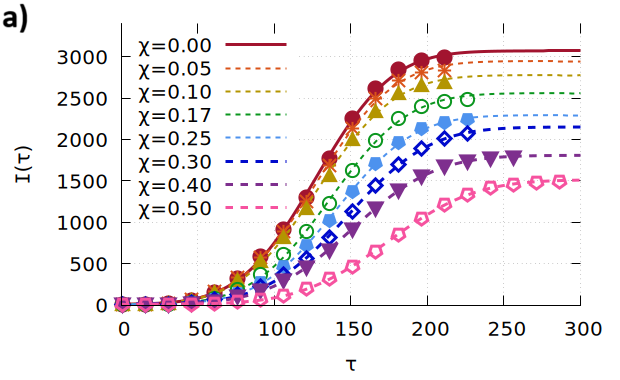}
\end{minipage}
\hfill
\begin{minipage}{0.3\textwidth}
    \centering
    \includegraphics[width=\linewidth]{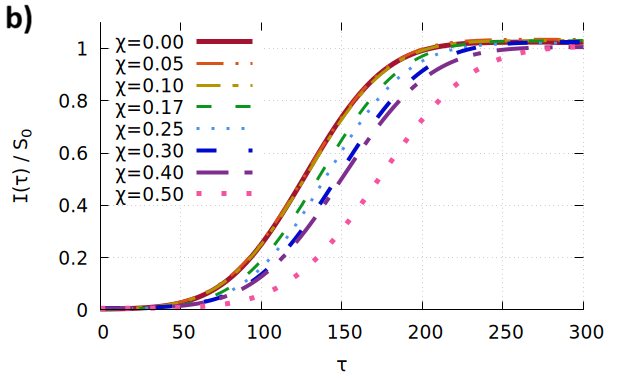}
\end{minipage}
\hfill
\begin{minipage}{0.3\textwidth}
    \centering
    \includegraphics[width=\linewidth]{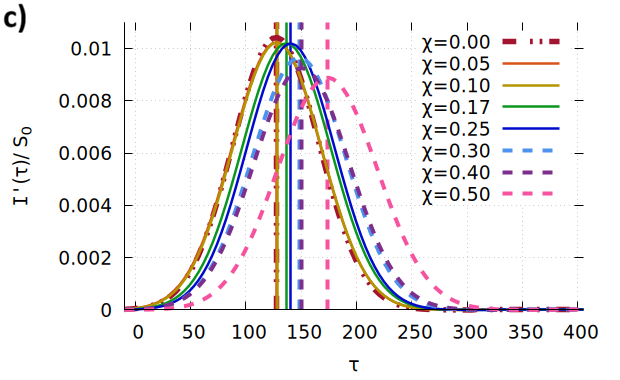}
\end{minipage}
\caption{\textbf{Results of the Packing fraction of 0.062}  (a) Contagion dynamics, $I(\tau)$; (b) normalised contagion dynamics, $I(\tau)/S_0$; and (c) normalised contagion rate, $I'(\tau)/S_0$. Symbols represent the averages of ten independent agent-based simulations, while the solid curves correspond to the fitted functions. Vertical lines in panel (c) indicate $\tau_{\mathrm{max}}$, the time at which the maximum contagion speed, $\nu_{\mathrm{max}}$, is reached.}
\end{figure}

\begin{figure}[ht]
\centering
\begin{minipage}{0.3\textwidth}
    \centering
    \includegraphics[width=\linewidth]{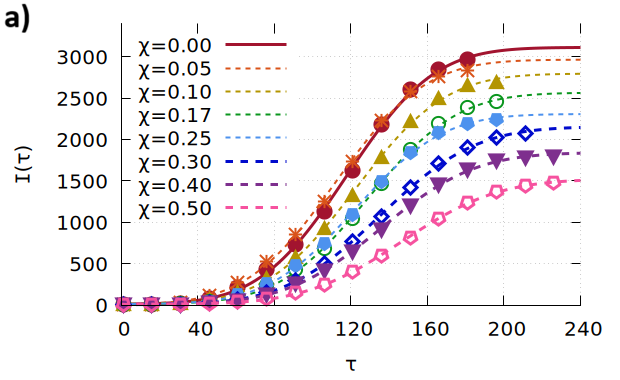}
\end{minipage}
\hfill
\begin{minipage}{0.3\textwidth}
    \centering
    \includegraphics[width=\linewidth]{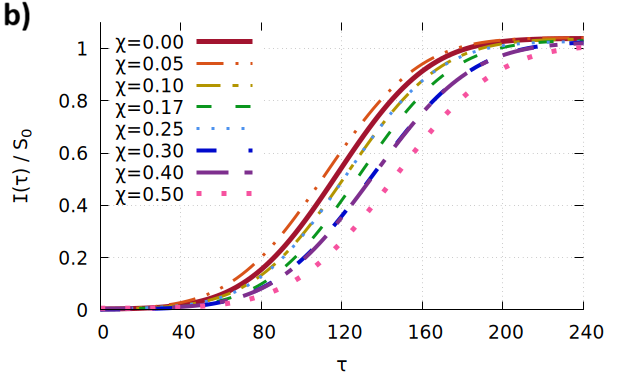}
\end{minipage}
\hfill
\begin{minipage}{0.3\textwidth}
    \centering
    \includegraphics[width=\linewidth]{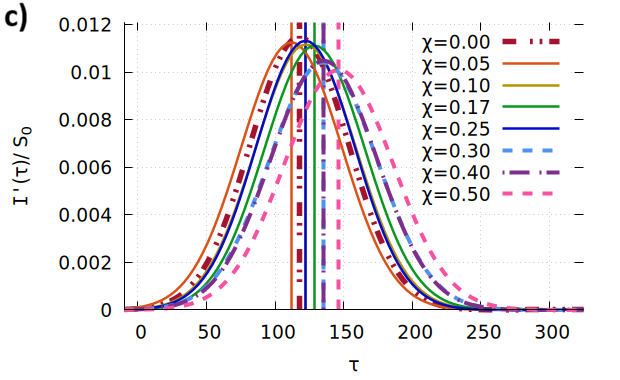}
\end{minipage}
\caption{\textbf{Results of the Packing fraction of 0.075}  (a) Contagion dynamics, $I(\tau)$; (b) normalised contagion dynamics, $I(\tau)/S_0$; and (c) normalised contagion rate, $I'(\tau)/S_0$. Symbols represent the averages of ten independent agent-based simulations, while the solid curves correspond to the fitted functions. Vertical lines in panel (c) indicate $\tau_{\mathrm{max}}$, the time at which the maximum contagion speed, $\nu_{\mathrm{max}}$, is reached.}
\end{figure}

\begin{figure}[ht]
\centering
\begin{minipage}{0.3\textwidth}
    \centering
    \includegraphics[width=\linewidth]{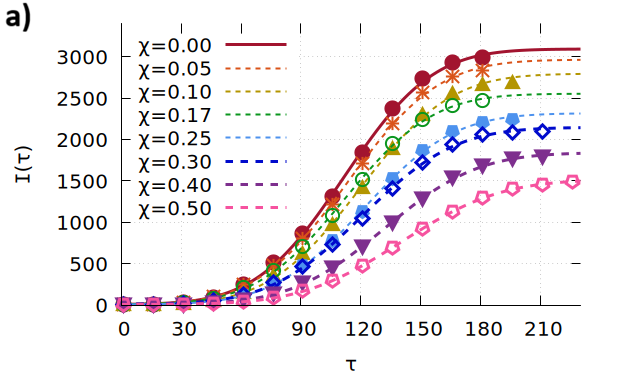}
\end{minipage}
\hfill
\begin{minipage}{0.3\textwidth}
    \centering
    \includegraphics[width=\linewidth]{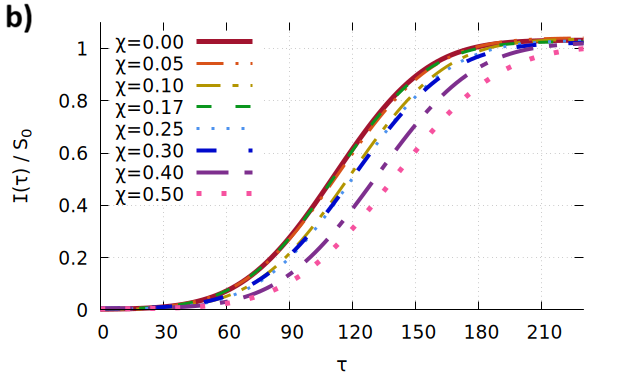}
\end{minipage}
\hfill
\begin{minipage}{0.3\textwidth}
    \centering
    \includegraphics[width=\linewidth]{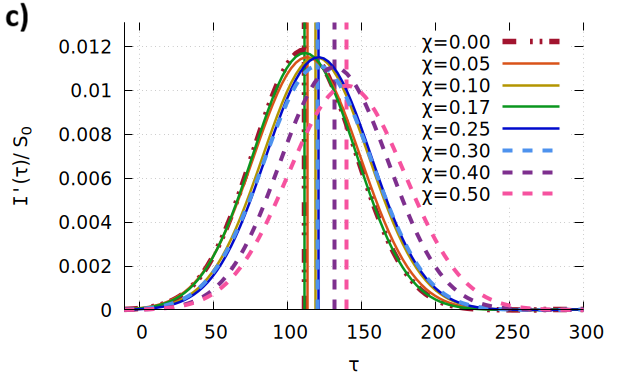}
\end{minipage}
\caption{\textbf{Results of the Packing fraction of 0.0813}  (a) Contagion dynamics, $I(\tau)$; (b) normalised contagion dynamics, $I(\tau)/S_0$; and (c) normalised contagion rate, $I'(\tau)/S_0$. Symbols represent the averages of ten independent agent-based simulations, while the solid curves correspond to the fitted functions. Vertical lines in panel (c) indicate $\tau_{\mathrm{max}}$, the time at which the maximum contagion speed, $\nu_{\mathrm{max}}$, is reached.}
\end{figure}

\begin{figure}[ht]
\centering
\begin{minipage}{0.3\textwidth}
    \centering
    \includegraphics[width=\linewidth]{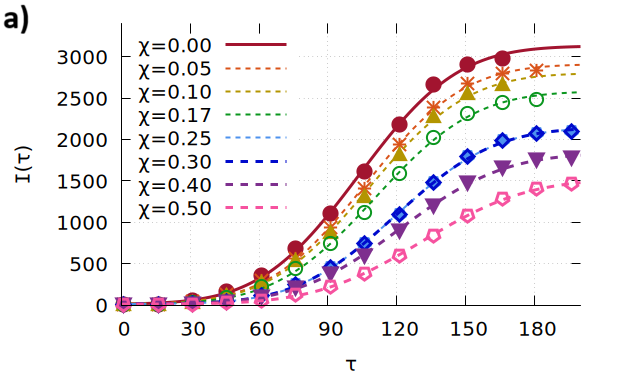}
\end{minipage}
\hfill
\begin{minipage}{0.3\textwidth}
    \centering
    \includegraphics[width=\linewidth]{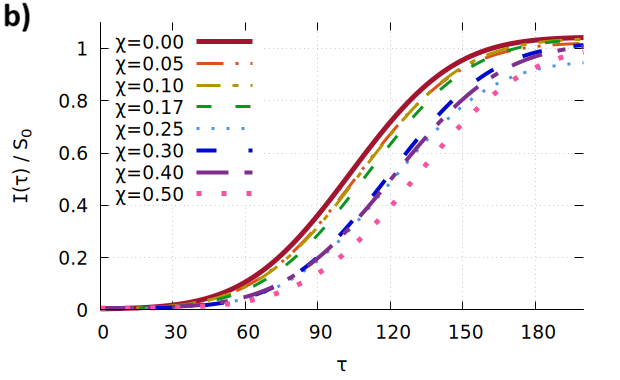}
\end{minipage}
\hfill
\begin{minipage}{0.3\textwidth}
    \centering
    \includegraphics[width=\linewidth]{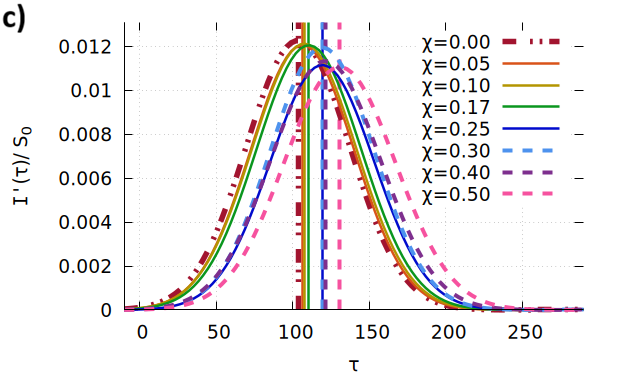}
\end{minipage}
\caption{\textbf{Results of the Packing fraction of 0.0875}  (a) Contagion dynamics, $I(\tau)$; (b) normalised contagion dynamics, $I(\tau)/S_0$; and (c) normalised contagion rate, $I'(\tau)/S_0$. Symbols represent the averages of ten independent agent-based simulations, while the solid curves correspond to the fitted functions. Vertical lines in panel (c) indicate $\tau_{\mathrm{max}}$, the time at which the maximum contagion speed, $\nu_{\mathrm{max}}$, is reached.}
\end{figure}

\begin{figure}[ht]
\centering
\begin{minipage}{0.3\textwidth}
    \centering
    \includegraphics[width=\linewidth]{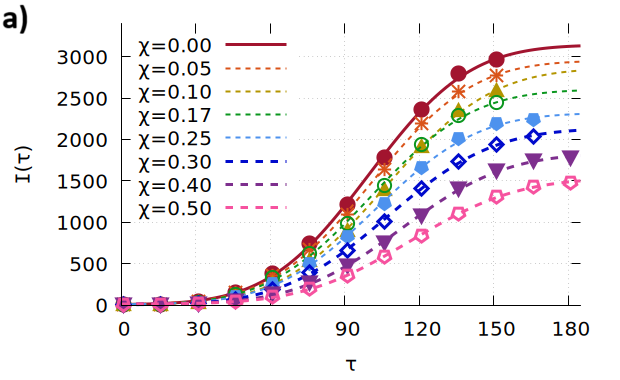}
\end{minipage}
\hfill
\begin{minipage}{0.3\textwidth}
    \centering
    \includegraphics[width=\linewidth]{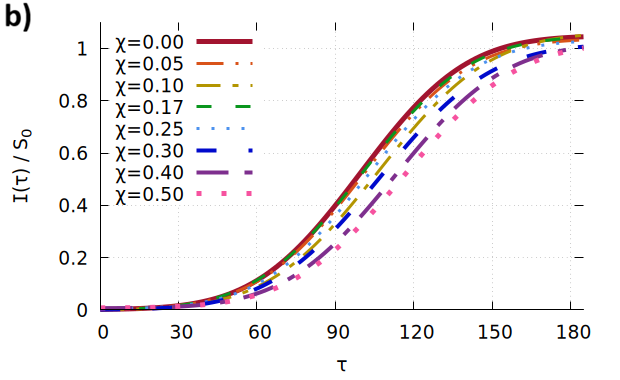}
\end{minipage}
\hfill
\begin{minipage}{0.3\textwidth}
    \centering
    \includegraphics[width=\linewidth]{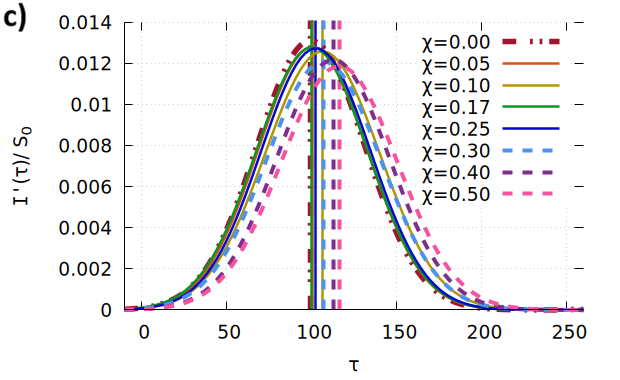}
\end{minipage}
\caption{\textbf{Results of the Packing fraction of 0.1}  (a) Contagion dynamics, $I(\tau)$; (b) normalised contagion dynamics, $I(\tau)/S_0$; and (c) normalised contagion rate, $I'(\tau)/S_0$. Symbols represent the averages of ten independent agent-based simulations, while the solid curves correspond to the fitted functions. Vertical lines in panel (c) indicate $\tau_{\mathrm{max}}$, the time at which the maximum contagion speed, $\nu_{\mathrm{max}}$, is reached.}
\end{figure}

\newpage
\section{normalised Maximum Contagion Rate for All Packing Fractions} \label{Appendix:2}

\begin{figure}[H]
\begin{center}
\includegraphics[width=0.6\linewidth]{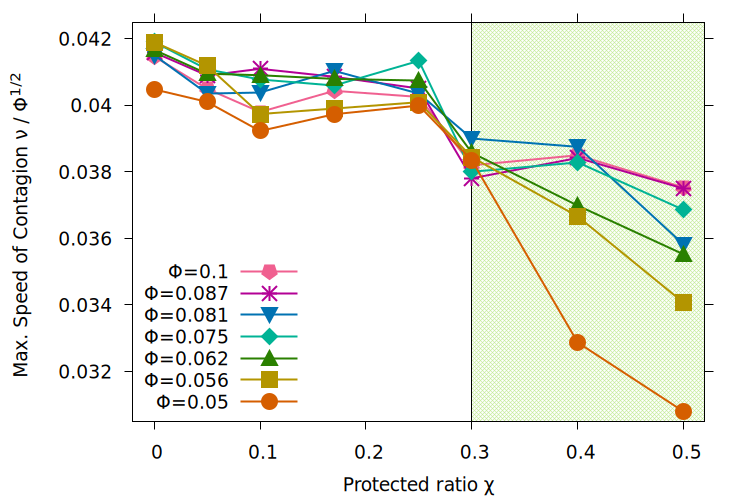} 
\caption{normalised maximum contagion rate, $\nu_{\mathrm{max}}\phi^{-1/2}$, as a function of the protected-agent fraction, $\chi$, for all packing fractions analysed in this work. Coloured symbols correspond to different values of $\phi$, and the solid lines are included as guides to the eye.}
\label{All_Speed_trim}
\end{center}
\end{figure}

This appendix presents the normalised maximum contagion rate, $\nu_{\mathrm{max}}\phi^{-1/2}$, as a function of the protected-agent fraction, $\chi$, for all packing fractions analysed in this work (Fig.~\ref{All_Speed_trim}).

A remarkable collapse of all curves is observed for $\chi < 0.3$, confirming that the maximum contagion rate follows the universal scaling $\nu_{\mathrm{max}}\propto\phi^{1/2}$, in the low-protection regime. 
This result indicates that, below the \textcolor{red}{crossover protected-agent fraction}, the contagion dynamics is primarily governed by the frequency encounter between agents, independently of the population density.

Above the \textcolor{red}{crossover}, $\chi \approx 0.3$, the curves progressively deviate from the universal scaling. Although the normalised maximum contagion rate decreases for all packing fractions, the reduction is considerably stronger in dilute systems than in dense ones. In particular, the highest packing fractions exhibit only a moderate decrease after the transition, whereas intermediate and low packing fractions show a pronounced suppression of the maximum speed of contagion as the protected-agent fraction increases.

\medline
\begin{multicols}{2}
\nocite{*}
\bibliographystyle{rmf-style}
\bibliography{References}

@article{Hu2013,
  author  = {Hu, Hao and Nigmatulina, Karima and Eckhoff, Philip A.},
  title   = {The scaling of contact rates with population density for the infectious disease models},
  journal = {Mathematical Biosciences},
  volume  = {244},
  number  = {2},
  pages   = {125--134},
  year    = {2013},
  doi     = {10.1016/j.mbs.2013.04.013}
}

@article{zhao_2024_farmed,
  author = {Zhao, Jin and (et. al) },
  month = {09},
  pages = {228–233},
  publisher = {Nature Portfolio},
  title = {Farmed fur animals harbour viruses with zoonotic spillover potential},
  doi = {10.1038/s41586-024-07901-3},
  url = {https://pubmed.ncbi.nlm.nih.gov/39232170/},
  year = {2024},
  journal = {Nature}
}

@misc{hantavirus,
  author = {{World Health Organization}},
  title = {Disease Outbreak News. Hantavirus cluster linked to cruise ship travel -- Multi-country.},
  year         = {2026},
  month        = may,
  day          = {4},
  howpublished = {\url{https://www.who.int/emergencies/disease-outbreak-news/item/2026-DON599}},
  note  = {Accessed June 16, 2026}
}

@misc{measles,
  author       = {{World Health Organization}},
  title        = {Disease Outbreak News: Measles in the Region of the Americas},
  year         = {2025},
  month        = apr,
  day          = {28},
  howpublished = {\url{https://www.who.int/emergencies/disease-outbreak-news/item/2025-DON565}},
  note         = {Accessed June 16, 2026}
}

@article{Mink2Humans,
author = {Bas B. Oude Munnink  and Reina S. Sikkema  and David F. Nieuwenhuijse  and Robert Jan Molenaar  and Emmanuelle Munger  and Richard Molenkamp  and Arco van der Spek  and Paulien Tolsma  and Ariene Rietveld  and Miranda Brouwer  and Noortje Bouwmeester-Vincken  and Frank Harders  and Renate Hakze-van der Honing  and Marjolein C. A. Wegdam-Blans  and Ruth J. Bouwstra  and Corine GeurtsvanKessel  and Annemiek A. van der Eijk  and Francisca C. Velkers  and Lidwien A. M. Smit  and Arjan Stegeman  and Wim H. M. van der Poel  and Marion P. G. Koopmans },
title = {Transmission of SARS-CoV-2 on mink farms between humans and mink and back to humans},
journal = {Science},
volume = {371},
number = {6525},
pages = {172-177},
year = {2021},
doi = {10.1126/science.abe5901},
URL = {https://www.science.org/doi/abs/10.1126/science.abe5901},
eprint = {https://www.science.org/doi/pdf/10.1126/science.abe5901}
}

@article{InfluenzA,
  title={Highly pathogenic avian influenza A (H5N1) virus in wild red foxes, the Netherlands, 2021},
  author={Rijks, Jolianne M and Hesselink, Hanna and Lollinga, Pim and Wesselman, Renee and Prins, Pier and Weesendorp, Eefke and Engelsma, Marc and Heutink, Rene and Harders, Frank and Kik, Marja and others},
  journal={Emerging infectious diseases},
  volume={27},
  number={11},
  pages={2960},
  year={2021}
}

@article {SIFrancisca,
	author = {Kermack, William Ogilvy  and McKendrick, A. G.  and Walker, Gilbert Thomas },
	title = {A contribution to the mathematical theory of epidemics},
	journal = {Proceedings of the Royal Society of London. Series A, Containing Papers of a Mathematical and Physical Character},
	volume = {115},
	number = {772},
	pages = {700-721},
	year = {1927},
	doi = {10.1098/rspa.1927.0118},
	URL = {https://royalsocietypublishing.org/doi/abs/10.1098/rspa.1927.0118},
	eprint = {https://royalsocietypublishing.org/doi/pdf/10.1098/rspa.1927.0118}
}

@article {Mallela,
	author = {Mallela, Abhishek},
	title = {Optimal Control applied to a SEIR model of 2019-nCoV with social distancing},
	elocation-id = {2020.04.10.20061069},
	year = {2020},
	doi = {10.1101/2020.04.10.20061069},
	publisher = {Cold Spring Harbor Laboratory Press},
	URL = {https://www.medrxiv.org/content/early/2020/04/22/2020.04.10.20061069},
	eprint = {https://www.medrxiv.org/content/early/2020/04/22/2020.04.10.20061069.full.pdf},
	journal = {medRxiv}
}

@misc {NanaSIR,
	doi = {10.48550/ARXIV.2004.11352},
	url = {https://arxiv.org/abs/2004.11352},
	author = {Bizet, Nana Cabo and de Oca, Alejandro Cabo Montes},
	title = {Modelos SIR modificados para la evoluci{\'o}n del COVID19},
	publisher = {arXiv},
	year = {2020},
	copyright = {arXiv.org perpetual, non-exclusive license}
}

@article {Waves,
	author = {Schwarzendahl, Fabian Jan and Grauer, Jens and Liebchen, Benno and L{\"o}wen, Hartmut},
	title = {Mutation induced infection waves in diseases like COVID-19},
	elocation-id = {2021.07.06.21260067},
	year = {2021},
	doi = {10.1101/2021.07.06.21260067},
	publisher = {Cold Spring Harbor Laboratory Press},
	URL = {https://www.medrxiv.org/content/early/2021/07/07/2021.07.06.21260067},
	eprint = {https://www.medrxiv.org/content/early/2021/07/07/2021.07.06.21260067.full.pdf},
	journal = {medRxiv}
}

@article {Saliva,
	doi = {10.1088/2399-6528/ac21a4},
	url = {https://doi.org/10.1088/2399-6528/ac21a4},
	year = 2021,
	month = {sep},
	publisher = {{IOP} Publishing},
	volume = {5},
	number = {9},
	pages = {095005},
	author = {A Calles and J L Mor{\'{a}}n-L{\'{o}}pez},
	title = {Modeling the viral load expelled in saliva droplets carrying {SARS}-{CoV}-2},
	journal = {Journal of Physics Communications}
}

@misc {Huepe,
	doi = {10.48550/ARXIV.2103.12618},
	url = {https://arxiv.org/abs/2103.12618},
	author = {Zhao, Yinong and Huepe, Cristi{\'{a}}n and Romanczuk, Pawel},
	title = {Contagion dynamics in self-organized systems of self-propelled agents},
	publisher = {arXiv},
	year = {2021},
	copyright = {Creative Commons Attribution 4.0 International}
}

@article {Pasillo,
	author = {Kramer,Kelby B.  and Wang,Gerald J. },
	title = {Social distancing slows down steady dynamics in pedestrian flows},
	journal = {Physics of Fluids},
	volume = {33},
	number = {10},
	pages = {103318},
	year = {2021},
	doi = {10.1063/5.0062331},
	URL = {https://doi.org/10.1063/5.0062331},
	eprint = {https://doi.org/10.1063/5.0062331}
}

@article {Levis,
	title = {Flocking-enhanced social contagion},
	author = {Levis, Demian and Diaz-Guilera, Albert and Pagonabarraga, Ignacio and Starnini, Michele},
	journal = {Phys. Rev. Research},
	volume = {2},
	issue = {3},
	pages = {032056},
	numpages = {7},
	year = {2020},
	month = {Sep},
	publisher = {American Physical Society},
	doi = {10.1103/PhysRevResearch.2.032056},
	url = {https://link.aps.org/doi/10.1103/PhysRevResearch.2.032056}
}

@article {Francisca,
	title = {Understanding contagion dynamics through microscopic processes in active Brownian particles},
	author = {Norambuena, Ariel and Valencia, Felipe J. and Guzm{\'{a}}n-Lastra, Francisca},
	journal = {Scientific Reports},
	volume = {10},
	issue = {1},
	pages = {20845},
	year = {2020},
	month = {Nov},
	doi = {10.1038/s41598-020-77860-y},
	url = {https://doi.org/10.1038/s41598-020-77860-y}
}

@article {AboutRobertBrown,
  author  = {Brian J. Ford},
  title   = {Brownian Movement in Clarkia Pollen: A Reprise of the First Observations},
  journal = {The Microscope},
  volume  = {40},
  number  = {4},
  pages   = {235--241},
  year    = {1992}
}

@article {Langevin1908,
	author = {Lemons, Don S.  and Gythiel, Anthony },
	title = {Paul Langevin's 1908 paper On the Theory of Brownian Motion [Sur la theorie du mouvement brownien C. R. Acad. Sci. (Paris) 146, 530 - 533 (1908)]},
	journal = {American Journal of Physics},
	volume = {65},
	number = {11},
	pages = {1079-1081},
	year = {1997},
	doi = {10.1119/1.18725},
	URL = {https://doi.org/10.1119/1.18725},
	eprint = {https://doi.org/10.1119/1.18725}
}

@book{Mazo2002,
  author    = {Robert M. Mazo},
  title     = {Brownian Motion: Fluctuations, Dynamics, and Applications},
  publisher = {Clarendon Press},
  address   = {Oxford},
  year      = {2002},
  series    = {International Series of Monographs on Physics},
  volume    = {112},
  isbn      = {9780198515678}
}

@book{PaperEinstein1905,
  author    = {Albert Einstein},
  title     = {Investigations on the Theory of the Brownian Movement},
  editor    = {R. Furth},
  publisher = {Dover Publications},
  address   = {New York},
  year      = {1956},
  series    = {Dover Books on Physics},
  isbn      = {9780486603049}
}

@article{Metareview2025,
  title={Metareview: a survey of active matter reviews},
  author={Te Vrugt, Michael and Wittkowski, Raphael},
  journal={The European Physical Journal E},
  volume={48},
  number={2},
  pages={12},
  year={2025},
  publisher={Springer}
}

@article{vicsek_model,
  title = {Novel Type of Phase Transition in a System of Self-Driven Particles},
  author = {Vicsek, Tam\'as and Czir\'ok, Andr\'as and Ben-Jacob, Eshel and Cohen, Inon and Shochet, Ofer},
  journal = {Phys. Rev. Lett.},
  volume = {75},
  issue = {6},
  pages = {1226--1229},
  numpages = {0},
  year = {1995},
  month = {Aug},
  publisher = {American Physical Society},
  doi = {10.1103/PhysRevLett.75.1226},
  url = {https://link.aps.org/doi/10.1103/PhysRevLett.75.1226}
}

@article{MIPS_paper,
	author = {Martin-Roca, Jos{\'{e}}  and Martinez, Raul  and Alexander, Lachlan C.  and Diez, Angel Luis  and Aarts, Dirk G. A. L.  and Alarc{\'{o}}n, Francisco  and Ram{\'{i}}rez, Jorge  and Valeriani, Chantal },
	title = {Characterization of MIPS in a suspension of repulsive active Brownian particles through dynamical features},
	journal = {The Journal of Chemical Physics},
	volume = {154},
	number = {16},
	pages = {164901},
	year = {2021},
	doi = {10.1063/5.0040141},
	URL = {https://doi.org/10.1063/5.0040141},
	eprint = {https://doi.org/10.1063/5.0040141}
}

@article{ermark1978,
	author = {Ermak,Donald L.  and McCammon,J. A. },
	title = {Brownian dynamics with hydrodynamic interactions},
	journal = {The Journal of Chemical Physics},
	volume = {69},
	number = {4},
	pages = {1352-1360},
	year = {1978},
	doi = {10.1063/1.436761},
	URL = {https://doi.org/10.1063/1.436761}
}

@book{Andrews1998,
  author    = {Larry C. Andrews},
  title     = {Special Functions of Mathematics for Engineers},
  publisher = {SPIE Press},
  address   = {Bellingham, WA},
  year      = {1998},
  isbn      = {9780819426167}
}

@article{MIPS_SIR_2022, 
	author={Forg{\'{a}}cs, P. and Lib{\'{a}}l, A. and Reichhardt, C. and Hengartner, N. and Reichhardt, C. J.},
	title={Using active matter to introduce spatial heterogeneity to the susceptible infected recovered model of epidemic spreading}, 
	volume={12}, 
	url={https://www.nature.com/articles/s41598-022-15223-5#citeas}, 
	DOI={10.1038/s41598-022-15223-5}, 
	number={1}, 
	journal={Scientific Reports}, 
	year={2022}
}

@Misc{gnuplot,
	author = { Williams, Thomas and Kelley, Colin and (et. all)},
	title = { Gnuplot 4.6: an interactive plotting program },
	month = { April },
	year = { 2013 },
	url = {http://gnuplot.sourceforge.net/}
}

@book{CodeC,
  author    = {Kernighan, Brian W. and Ritchie, Dennis M.},
  title     = {The C Programming Language},
  edition   = {2nd},
  year      = {1988},
  publisher = {Prentice Hall},
  address   = {Englewood Cliffs, NJ},
  isbn      = {0-13-110362-8}
}

@book{numericalrec,
    author    = "Press, W. H. and Teukolsky, S. A. and Vetterling, W. T. and Flannery, B. P.",
    title     = "Numerical Recipes in C: The Art of Scientific Computing",
    edition={Third},
    year      = "2007",
    publisher = "Cambridge University Press",
    address   = "Cambridge"
}

@article{Kenah_2011,
author = {Kenah, Eben and Miller, Joel C.},
title = {Epidemic Percolation Networks, Epidemic Outcomes, and Interventions},
journal = {Interdisciplinary Perspectives on Infectious Diseases},
volume = {2011},
number = {1},
pages = {543520},
doi = {https://doi.org/10.1155/2011/543520},
url = {https://onlinelibrary.wiley.com/doi/abs/10.1155/2011/543520},
eprint = {https://onlinelibrary.wiley.com/doi/pdf/10.1155/2011/543520},
year = {2011}
}

@article{Yang_2015,
   title={Large epidemic thresholds emerge in heterogeneous networks of heterogeneous nodes},
   volume={5},
   issue= {1},
   ISSN={2045-2322},
   url={https://doi.org/10.1038/srep13122},
   DOI={10.1038/srep13122},
   number={5},
   journal={Scientific Reports},
   author={Yang, Hui and Tang, Ming and Gross, Thilo},
   year={2015},
   month= 08 }

@article{Hiraoka_2022,
   title={Herd immunity and epidemic size in networks with vaccination homophily},
   volume={105},
   ISSN={2470-0053},
   url={http://dx.doi.org/10.1103/PhysRevE.105.L052301},
   DOI={10.1103/physreve.105.l052301},
   number={5},
   journal={Physical Review E},
   publisher={American Physical Society (APS)},
   author={Hiraoka, Takayuki and Rizi, Abbas K. and Kivelä, Mikko and Saramäki, Jari},
   year={2022},
   month=May }

@article{Review_2016,
	doi = {10.1103/revmodphys.88.045006},
	url = {https://doi.org/10.1103%2Frevmodphys.88.045006},
	year = 2016,
	month = {nov},
	publisher = {American Physical Society ({APS})},
	volume = {88},
	number = {4},
	author = {Clemens Bechinger and Roberto Di Leonardo and Hartmut L{\"{o}}wen and Charles Reichhardt and Giorgio Volpe and Giovanni Volpe},
	title = {Active Particles in Complex and Crowded Environments},
	journal = {Reviews of Modern Physics}
}

@Article{ABP_Joakin2014,
	author ={Stenhammar, Joakim and Marenduzzo, Davide and Allen, Rosalind J. and Cates, Michael E.},
	title  ={Phase behaviour of active Brownian particles: the role of dimensionality},
	journal  ={Soft Matter},
	year  ={2014},
	volume  ={10},
	issue  ={10},
	pages  ={1489-1499},
	publisher  ={The Royal Society of Chemistry},
	doi  ={10.1039/C3SM52813H},
	url  ={http://dx.doi.org/10.1039/C3SM52813H}
}

@article{peruani_sibona_2008, 
	author={Peruani, Fernando and Sibona, Gustavo J.},	
	title={Dynamics and steady states in Excitable Mobile Agent Systems}, 
	volume={100}, 
	url={https://pubmed.ncbi.nlm.nih.gov/18518251/}, 
	DOI={10.1103/physrevlett.100.168103}, 
	number={16}, 
	journal={Physical Review Letters},  
	year={2008}
}

@article{Ginot2018,
  author  = {Ginot, F. and Theurkauff, I. and Detcheverry, F. and Ybert, C. and Cottin-Bizonne, C.},
  title   = {Aggregation-fragmentation and individual dynamics of active clusters},
  journal = {Nature Communications},
  year    = {2018},
  volume  = {9},
  number  = {1},
  pages   = {696},
  doi     = {10.1038/s41467-017-02625-7}
}

@article{Alarcon2017,
  author  = {Alarc{\'o}n, Francisco and Valeriani, Chantal and Pagonabarraga, Ignacio},
  title   = {Morphology of clusters of attractive dry and wet self-propelled spherical particle suspensions},
  journal = {Soft Matter},
  year    = {2017},
  volume  = {13},
  number  = {4},
  pages   = {814--826},
  doi     = {10.1039/C6SM01752E}
}

@misc{simulation_Videos,
  author       = {Regalado, I. Sicarú and Alarcón, Francisco},
  title        = {Simulation Videos},
  year         = {2024},
  howpublished = {\url{https://canva.link/pqdv9m6x8aemu52}},
  note         = {Supplementary simulation videos Contagion Dynamics}
}
\end{multicols}
\end{document}